\documentstyle[aps,multicol,epsfig,eqsecnum]{revtex}
\newcommand\be{\begin{eqnarray}}
\newcommand\ee{\end{eqnarray}}
\newcommand\ba{\begin{array}}
\newcommand\ea{\end{array}}
\def\r{\rangle}
\def\l{\langle}
\def\T{{\rm Tr}}
\begin{document}
\title{
On the local unitary equivalence of states of 
multi-partite systems}
\author{
M. Ziman${}^{1}$, P. \v{S}telmachovi\v{c}${}^{1}$, and
V. Bu\v{z}ek${}^{1,2}$
}
\address{
${}^{1}$ Research Center for Quantum Information,
Slovak Academy of Sciences,
D\'ubravsk\'a cesta 9, 842 28 Bratislava, Slovakia \\
${}^{2}$Faculty of Informatics, Masaryk University, Botanick\'a 68a,
602 00 Brno, Czech Republic 
}
\date{25 June 2001}
\maketitle
\begin{abstract}
Two pure states of a multi-partite system are alway are related
by a unitary transformation acting on the Hilbert space of the
whole system. This transformation involves multi-partite transformations.
On the other hand some quantum information protocols such as
the quantum teleportation and quantum dense coding are based
on equivalence of some classes of states of bi-partite systems 
 under the action of  local (one-particle) unitary operations. 
In this paper we address the question: ``Under what
conditions are the two states 
states, $\varrho$ and $\sigma$, of a multi-partite system
locally unitary equivalent?'' We present a set of conditions
which have to be satisfied in order that the two states are locally
unitary equivalent. In addition, we study whether
it is possible to prepare a state of 
a multi-qudit system. which is 
divided into two parts $A$ and $B$, by  unitary operations acting only
on the systems $A$ and $B$, separately.
\end{abstract}
\begin{multicols}{2}
\section{Introduction}
Quantum entanglement is the key ingredient in  quantum information
processing \cite{Nielsen2000}.
Bipartite entanglement is quite ``well'' understood by now,
but the investigation of the multi-partite case is just  beginning.

Several 
quantum information protocols such as
quantum teleportation and quantum dense coding are based
on the equivalence of some classes of states of bi-partite systems
 under the action of
local (one-particle) unitary operations \cite{Nielsen2000,Gruska1999}.
So the question of local unitary equivalence of multi-partite states of 
quantum systems is of importance.
Recently several authors 
\cite{Werner00,Sudbery01,Barnum01} have 
studied the so-called {\it polynomial invariants} of local unitaries.
In the present paper we discuss the problem of  local unitary
equivalence. We present a set of necessary conditions which the two 
multi-partite states have to fulfill in order to declare them
{\em locally unitary equivalent}.

The state of a quantum system is usually represented 
by  a {\it density matrix.}
We know that the elements of the density matrix depend on the choice
of the basis in Hilbert space, $\cal{H}$, of the system.
Physically the basis can be thought of as a   
{\it complete set of observables}. 
For example, in the case
of spin-1/2 particle, the choice of the basis in Hilbert space
corresponds to the direction of the Stern-Gerlach experimental setup
\cite{Peres1995}. 

A local unitary transformation corresponds to a change of a basis
in each of the subsystems. 
Such unitary transformations are known as  {\it passive} transformations, 
because there is no physical action performed on the quantum system
itself. These transformations simply reflect the choice of our point of view
rather than 
 any specific manipulation of the physical system.
It reflects the fact that two locally-unitary-equivalent states
have the same matrix form, only the choice basis of subsystems is different.
An {\it active} unitary transformation corresponds to the most general
dynamical map of the states for isolated systems. It means that
 under the dynamics
described by local unitaries each of the subsystems evolves independently.
Physically it corresponds to the absence of interactions
between the subsystems.
This property allows us to prepare from
a given 
state of the multi-partite system  some other state
only by using {\em local} unitaries (here we assume 
that the subsystems can be separated by space-like intervals).
From a mathematical point of view there is no difference between 
the active and passive unitary transformations.

In this paper we address the question: ``Under what conditions are
 the two states 
 states, $\varrho$ and $\sigma$, of a multi-partite system
 locally unitary equivalent?'' We present a set of conditions
which have to be satisfied in order that the two states are locally
unitary equivalent. In addition we study whether
it is possible to prepare a state of 
a multi-qudit system which is 
divided into two parts $A$ and $B$ by  unitaries acting only
on the systems $A$ and $B$, separately.
In Section \ref{sec2} 
we present examples of some invariants together with some 
applications for pure states. A criterion about the local unitary 
equivalence for mixed states is given in Section
\ref{sec3}.
In Section \ref{sec4} we investigate  
the connection of bipartite and multi-partite 
entanglement via local unitaries for system of $N$ qudits. 

\section{Local unitary equivalence}
\label{sec2}
Let us 
consider two states $\varrho$ and $\sigma$
of an $N-$party system described by the Hilbert space
${\cal{H}}_{A_1\dots A_N}$.
The question of local unitary equivalence can be mathematically formulated 
in the following way: 
{\it Does there exist a unitary transformation
$U=U_1\otimes\dots\otimes U_N$, such that $\varrho=U\sigma U^\dagger$?}

In order to study this problem of local unitary equivalence we 
firstly, simplify our task. Specifically, we divide the total Hilbert space 
in the following way
\be
{\cal{H}}_{A_1\dots A_N}={\cal{H}}_{A_1}\otimes\dots\otimes{\cal{H}}_{A_N}=
{\cal{H}}_X\otimes{\cal{H}}_Y.
\label{2.1}
\ee
where $X$ and $Y$ are two sets  of subsystems $A_k$.
If we would have some criteria for bi-partite systems being locally unitary 
equivalent, then we can decide whether
the following is true
\be
\varrho=U_X\otimes U_Y\sigma U_X^\dagger\otimes U_Y^\dagger .
\label{2.2}
\ee
If the answer is ``no'', then the states are not locally unitary equivalent.
If the answer is ``yes'', then we continue our investigation by 
checking whether the pair of states
\be
\varrho_X=\T_Y\varrho & \ \ {\rm and} \ \ & \sigma_X=\T_Y\sigma 
\nonumber
\\
\varrho_Y=\T_X\varrho & \ \ {\rm and} \ \ & \sigma_Y=\T_X\sigma
\label{2.3}
\ee
are locally unitary equivalent. That is, we 
again, divide each of the subsystems $X,Y$ 
into two smaller sets and we use the criteria for bi-partite system.
Using this method recursively we finally  partition the multi-partite
system into its components 
 $A_k$ and give an answer to the question of locally unitary 
equivalence. From here 
it follows that it is enough to have a criterion for the bi-partite
composite systems with the help of which multi-partite systems can
be studied. In what follows , we present some necessary conditions for 
states to be locally unitary equivalent. 
\subsection{Invariants of local unitaries}
\label{subsec2.1}
\begin{enumerate}
\item{\bf Eigenvalues.}
It is well known that the necessary and sufficient condition for 
two states to be unitarily equivalent is that
they have the same set of eigenvalues. Let us 
denote by $eig(\varrho)$
the set of eigenvalues of the density matrix $\varrho$. Then 
\be
\varrho^\prime =U\varrho U^\dagger,\ \ {\rm iff}\ \ 
eig(\varrho^\prime)\equiv eig(\varrho).
\label{2.4}
\ee
\item{\bf Eigenvalues of the reduced states.}
The reduced density matrices $\varrho_X$ and $\varrho_Y$ change
after the action of 
the local unitary transformation $U_X\otimes U_Y$ as follows:
for all $U_X$ and $U_Y$ 
\be
\varrho^\prime_X &=& \T_Y \left[ U_X\otimes U_Y\varrho 
U^\dagger_X\otimes U^\dagger_Y \right]=U_X \varrho_X U_X^\dagger \, ,
\nonumber
\\
\varrho_Y^\prime &=& \T_X \left[ U_X\otimes U_Y\varrho 
U^\dagger_X\otimes U^\dagger_Y \right]=U_Y \varrho_Y U_Y^\dagger \, .
\label{2.5}
\ee

Now we can define 
the first invariant:  local unitary transformations
do not change the eigenvalues of the reduced density matrices,
which describe states of the subsystems.

For any  state $\varrho^\prime$ such that
$\varrho^\prime =U_X\otimes U_Y\varrho U^\dagger_X\otimes U^\dagger_Y$
it is valid that 
\be
\nonumber
 eig(\varrho_X) &\equiv& eig(\varrho^\prime_X) \, ,\\ 
eig(\varrho_Y) & \equiv& eig(\varrho^\prime_Y)\, .
\label{2.6}
\ee
We note that this relation is valid for any operator $\varrho$,
not only for density operators.

\item{\bf Eigenvalues of the reduced eigenprojectors.}
Quantum 
states can be expressed in a a  spectral form. Specifically,  
let $U=U_X\otimes U_Y$ and
$\varrho=\sum_k \lambda_k P_k$, then $\varrho^\prime=
U\varrho U^\dagger=\sum_k \lambda_k P^\prime_k$,
where $P_k$ and $P^\prime_k=UP_kU^\dagger$ are eigenprojectors belonging to 
the eigenvalue $\lambda_k$. 
Since the unitary transformation under consideration is local, 
i.e.  $U=U_X\otimes U_Y$,
we obtain the implication
\be
\nonumber
\varrho^\prime=U\varrho U^\dagger & 
\ \ {\rm then}\ \ & dim(P_k)=dim(P^\prime_k)\\
\nonumber & & eig(\T_Y P_k)\equiv eig(\T_Y P^\prime_k)\\
 & & eig(\T_X P_k)\equiv eig(\T_X P^\prime_k)\, ,
\label{2.7}
\ee
where $dim(P_k)$ denotes the dimension of the projection $P_k$ 
defined as the number of its nonzero eigenvalues.

\item{\bf Partial transposition.}
We use the following identity for any operators
$K_X,L_Y,M_X,N_Y$
\be
(K_X\otimes L_Y\varrho M_X\otimes N_Y )^{T_Y}=K_X\otimes N_Y^T\varrho^{T_Y}
M_X\otimes L_Y^T
\label{2.8}
\ee
where by $T_Y$ we denote the partial transposition of the subsystem $Y$.
A similar identity is fulfilled also for the partial transposition, 
$T_X$,  of the system $X$. 
Note, that the partial transposition depends on the choice of the local basis.
Introducing the local unitary transformation $U_X\otimes U_Y$
we obtain
\be
\nonumber
(\varrho^\prime )^{T_Y} &=&
(U_X\otimes U_Y\varrho U_X^\dagger\otimes U_Y^\dagger )^{T_Y}\\ &=&
U_X\otimes (U_Y^\dagger)^T\varrho^{T_Y}U_X^\dagger\otimes U_Y^T 
\nonumber
\\ 
 &=&
W\varrho^{T_Y}W^\dagger
\label{2.9}
\ee
where $W=U_X\otimes (U_Y^T)^\dagger$ is a unitary operator. 
Its unitarity follows from   the fact that the  
transposed unitary operator is a unitary operator. 
The unitary transformation
leaves the eigenvalues of the operator $\varrho^{T_Y}$ unchanged. Therefore, 
we obtain the second invariant associated with  density matrices 
which are invariant under local
unitary transformations. Namely, 
if $\varrho^\prime =U_X\otimes U_Y\varrho U_X^\dagger\otimes U_Y^\dagger $
then 
\be
\nonumber
eig(\varrho^{T_Y})&\equiv & eig(\varrho^{\prime T_Y})\\ 
eig(\varrho^{T_X})&\equiv & eig(\varrho^{\prime T_X}).
\label{2.10}
\ee 
\end{enumerate}
\subsection{Local unitary equivalence for pure states}
\label{subsec2.2}
Let us consider 
two pure states $|\psi\r$ and $|\phi\r$ of a bi-partite system 
written in the Schmidt-decomposed form
\be
\nonumber
|\psi\r &=& \sum_k \sqrt{\lambda_k} |k\r_X\otimes|k\r_Y \\
|\phi\r &=& \sum_k \sqrt{\mu_k} |k^\prime\r_X\otimes|k^\prime\r_Y .
\label{2.11}
\ee
From the equality $\lambda_k=\mu_k$ (for all $k$) the validity of the 
second invariant property follows. That is 
 the reduced states have the same set 
of eigenvalues, namely $\{\lambda_k\}$. The first invariant property is 
fulfilled trivially for pure states. Moreover, 
let us define  local unitaries
by the equations $U_X|k\r_X=|k^\prime\r_X$ and $U_Y|k\r_Y=|k^\prime\r_Y$.
Then  the states (\ref{2.11}) 
are locally unitary equivalent and $|\phi\r=U_X\otimes U_Y
|\psi\r$.

{\bf Theorem 1}\\
{\it Let us consider 
two pure states 
 $|\psi\r$ and $|\phi\r$ of a bi-partite system.
These state are locally unitary equivalent if and only if their coefficients
in the Schmidt decomposition are equal, i.e. the sets of eigenvalues
of reduced states coincide.} 

{\em Example 1.}\newline
This simple theorem can be rather useful. In particular, we can use it to
prove that 
any GHZ state is locally equivalent to
one EPR pair and some ancilla system in an arbitrary state. It means we
are able to prepare a GHZ state from an
 EPR pair a vice versa \cite{Nielsen2000}.
Let us define two states of $N$ qubits 
$|GHZ\r_{1\dots N}=(|0^{\otimes(N)}\r+|1^{\otimes(N)}\r)/\sqrt{2}$
and $|\Psi\r_{1\dots N}=|EPR\r_{12}\otimes|\phi\r_{3\dots N}$, where
$|EPR\r=(|00\r+|11\r)/\sqrt{2}$. Since $\T_{2\dots N}|GHZ\r\l GHZ|=
\T_{2\dots N}|\Psi\r\l\Psi|=\frac{1}{2}\openone$ we know that the eigenvalues
of the reduced states coincide. Specifically, let us consider a splitting
of the whole system into a single qubit denoted as ``1'' and the rest of
the system. It means that 
these states are locally (with respect to the given splitting)
 unitary equivalent.

{\em Example 2.}\newline
Let us consider a  state (for more details see 
\cite{Stelmachovic01})
\be
|X\r=\frac{1}{\sqrt{2^n}}\sum_{k=0}^{2^n-1}
|k\r_X\otimes |\alpha_k\r_Y
\label{2.12}
\ee  
of  $N=2n+1$ qubits located 
on a ring. This system is divided 
into two subsystem of
$n$ neighboring qubits (subsystem $X$) and $(n+1)$ neighboring qubits
(subsystem $Y$). 
In this notation $|k\r_X$ represents the state of $n$ qubits, 
if $k$ is written in a binary form. The state $|\alpha_k\r_Y=\sum_l 
(-1)^{d_k+d_l+d_{k,l}}|l\r_Y$ where $|l\r_Y$ is the state of the rest
of the 
$n+1$ qubits, if $l$ is written in a binary form. The number $k$ and $l$
define the state of the ring in the sense of its binary expression. From
here we can determine the state of each individual qubit.
The numbers $k$ give us the qubits in the state $|1\r$ and we can write
$\{i,\dots,j\}$ to represents the position of the qubits in this state.
Similarly,  $l$ determines the set $\{r,\dots,s\}$.
We define  $d_k=d_{\{i,\dots,j\}}:=\sum_{i<j}d(i,j)$,
$d_{k,l}=\sum_{a\in \{i,\dots j\},b\in\{r,\dots,s\}}d(a,b)$, 
where $d(i,j)$ is the shortest 
distance between two qubits (in the state $|1\r$) on a ring (i.e.
number of steps needed to get from qubit in position $i$ into the position
$j$ around the ring, or the smallest number of qubits lying between 
the $i$-th and $j$-th position). 
The  state (\ref{2.12}) is one of the eigenstates of the 
Ising model (for more details see \cite{Stelmachovic01}).

For the purpose of our example 
it is important that  Eq.~(\ref{2.12}) represents
the Schmidt form of the state $|X\r$ with all Schmidt coefficients 
equal to $1/\sqrt{2^n}$ and the Schmidt number (i.e. number of nonzero
Schmidt coefficients) is equal to $2^n$. Let us define a state
\be
\label{2.13}
|\Psi\r=|\phi\r_{Y_{n+1}}
\otimes\left(\bigotimes_{j=1}^n|EPR\r_{X_jY_j}\right)\, ,
\ee
where $|EPR\r_{X_jY_j}=(|0\r_{X_j}\otimes |0\r_{Y_j}+|1\r_{X_j}\otimes
|1\r_{Y_j})/\sqrt{2}$ is the EPR pair shared between the subsystems $X$ and
$Y$, such that $j$ represents the position of the qubit in this subsystems.
We can express the state (\ref{2.13}) in the  Schmidt form
\be
\label{2.14}
|\Psi\r=\frac{1}{\sqrt{2^n}}\sum_{k=0}^{2^n-1}|k\r_X\otimes|\beta_k\r_Y \, ,
\ee
where $|\beta_k\r_Y=|k\r_{Y^\prime}\otimes|\phi\r_{n+1}$ and $Y^\prime$
denotes the subsystem $Y$ without one of its qubit at the $(n+1)$-th position.
Since the set of Schmidt coefficients for these states coincides, we 
conclude that the states
 (\ref{2.12}) and (\ref{2.13})
are locally unitarily equivalent. We stress that 
this equivalence is with respect to the splitting of the ring into two
parts $X$ and $Y$.
Therefore, 
from the point of view of entanglement between systems $X$
and $Y$  these states are equivalent, and we are able to prepare $n$ EPR
pairs between these systems by a 
local transformation $U=\openone_X\otimes U_Y$,
where $U_Y$ is determined by the equality $U_Y|\alpha_k\r_Y=|\beta_k\r_Y$. 
Finally we note, 
that this transformation is not uniquely defined, because of the 
degeneracy of Schmidt coefficients.

\subsection{Counterexample for mixed states}
\label{subsec2.3}

Unfortunately,  Theorem 1 is valid only for pure states
and cannot be generalized for arbitrary impure
 states. 
In what follows we will show a simple counterexample. 
The idea behind this example
is that we cannot create entanglement by a local unitary 
transformation and that the spectral decomposition of a density matrix
is unique. It means the separable pure state in the spectral decomposition 
cannot evolve into entangled one.

{\em Example 3.}\newline
Let us 
consider two states $\varrho_{XY}$ and $\sigma_{XY}$ of a bi-partite system
$\cal{H}_X\otimes \cal{H}_Y$, where $dim{\cal H}_Y=2$ and $dim{\cal H}_Y=8$.
This means the system of four qubits
is  divided into a one-qubit and a three-qubit subsystems.
Let the set of eigenvalues be the same
\be
eig(\varrho_{XY})=eig(\sigma_{XY})=
\left\{\frac{1}{4},\frac{3}{8},\frac{5}{16},
\frac{1}{16}\right\}
\label{2.15}
\ee
and the corresponding eigenvectors be 
\be
\label{2.16}
\left(
\begin{array}{c} 
\frac{1}{\sqrt{4}} 
\left(|0,1\r +|0,2\r + |0,4\r+|1,0\r\right) \\
|1,7\rangle
\\
|0,5\r  \\
\frac{1}{\sqrt{3}}(|1,1\r+|1,2\r+|1,4\r) 
\end{array}
\right)
\ee
for $\varrho_{XY}$ and 
\be
\label{2.17}
\left(\begin{array}{c}
|1,7\r
\\
\frac{1}{\sqrt{3}}(|0,7\r+|1,3\r +|0,5\r)
\\
|0,0\r \\
|0,1\r
\end{array}
\right)
\ee
for $\sigma_{XY}$.
It is easy to check that these two states also
have  the same sets of 
eigenvalues of their  reduced density matrices, i.e.
\be
\nonumber
eig(\varrho_X)& = & eig(\sigma_X)=\left\{ \frac{1}{2},\frac{1}{2}\right\} 
\\
eig(\varrho_Y)& = & eig(\sigma_Y)=\left\{ \frac{3}{8},\frac{5}{16},
\frac{1}{4},\frac{1}{16}\right\}
\label{2.19}
\ee
So for the state under considerations 
the conditions of  Theorem 1 are satisfied. On the other hand, 
in contradiction to the general properties of local unitaries,
we see the violation of 
the  property 
the entanglement creation prohibited 
by local unitary transformation.  
In other words, the condition of the same set
of reduced eigenvalues of eigenprojectors (invariant A3) is not fulfilled.

\section{Local unitary equivalence for the mixed states}
\label{sec3}
We have shown that  Theorem 1 is not valid for all  states.
Certainly, 
it can be trivially generalized to a class of the so called
 {\it factorizable states},
given by density operators $\varrho_{XY}=\varrho_X\otimes\varrho_Y$.
On the other hand,
it is not clear whether  the invariants presented in Section
\ref{sec2} 
provide us with  a definite answer for the problem of 
the local unitary  equivalence for mixed states.
To clarify the situation we 
formulate the following theorem\\

{\bf Theorem 2}\\
{\it Suppose two non-degenerate states $\varrho$ and $\sigma$ of the composite 
system $X+Y$ are given,  i.e.
\be
\varrho=\sum_k\lambda_k|\psi_k\r\l\psi_k| \, , \qquad 
\sigma=\sum_k\mu_k|\phi_k\r\l\phi_k|.
\ee
and $\lambda_k\ne\lambda_l,\mu_k\ne\mu_l$ for $k\ne l$. 
Let us express the vectors $|\psi_1\r$ and $|\phi_1\r$ in their Schmidt bases
and fix these two basis on ${\cal H}_{X}\otimes{\cal H}_Y$.
If for each $k$, $\lambda_k=\mu_k$  and the coefficients of the corresponding
eigenvectors 
$|\psi_k\r=\sum_{mn}\alpha^k_{mn}|mn\r$ and 
$|\phi_k\r=\sum_{mn}\beta^k_{mn}|m^\prime n^\prime\r$ 
(where $|mn\r$ and $|m^\prime n^\prime\r$ are the previously 
mentioned fixed bases)
coincide, i.e. $\alpha^k_{mn}=\beta^k_{mn}$, 
then the states $\varrho$, $\sigma$ 
are locally unitary equivalent. The local transformation is given by
$U_X|m\r=|m^\prime\r$ and $U_Y|n\r=|n^\prime\r$.}
 
From the construction it is clear that the theorem is valid. Note that
the choice of the fixed basis is given by the Schmidt expression of
one of the eigenvectors, but the result doesn't depend on it. In the 
degenerate case we cannot use the argument with the Schmidt decomposition,
because then the eigenprojectors are no longer one dimensional. 

We conclude this section by saying that 
in general by using invariants of local unitaries 
we are able only to determine whether 
the states under consideration are not locally unitary equivalent.

\section{Local connection of multi-partite and bipartite entanglement}
\label{sec4}
\begin{figure}[h]
\centerline{\epsfig{width=8.0cm, file=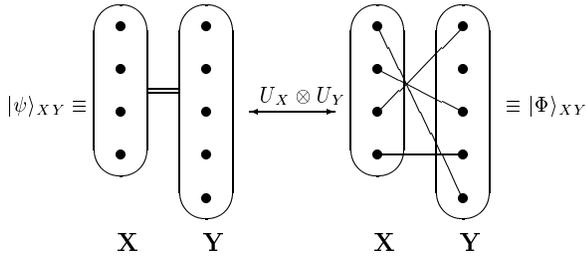}} 
\caption{Schematic representation of
a  connection
between multi-partite and bipartite entanglement
via local operations. On the left side
of the figure the double line represents the shared multi-partite
entanglement between $X$ and $Y$.
Each line in the right side of the figure 
represents an entangled pair of qudits 
$|\chi\r_j$, i.e. bipartite entanglement.}
\label{fig1}
\end{figure}

In this section we generalize  Example 2 of  Section \ref{sec3}.
Specifically, let us consider 
a pure state $|\psi\r_{A_1,\dots,A_N}$ of a composite system consisting
of $N$ qudits. Let us split this system into 
two multi-partite subsystems $X$ and $Y$.
By calculating the entropy of one of the subsystems we can determine  
the degree 
of  entanglement between these two parties. An interesting question
is whether with the help of a local transformation with respect to the
splitting we are able to
 prepare a state where the entanglement is shared only between
the specific pairs of 
qudits such that one particle of each pair belongs to the system $X$
while the second belongs to the system $Y$. In other words, the
qudits 
belonging to the subsystem $X$ ($Y$) are mutually separable. 
It means that the multi-partite entanglement shared between the
subsystems $X$ and $Y$ can be arranged by bi-partite entanglement shared
between the pairs of qudits  just by performing
an appropriate local unitary transformation $U=U_X\otimes U_Y$
(see Fig.~\ref{fig1}).

Let us denote 
 by $A_1,\dots,A_N$ the qudits of the whole system and by $X_1,\dots,X_n$
($Y_1,\dots,Y_m$)
the qudits  in a subsystem $X$ ($Y$).  Note that $n+m=N$. 
We also introduce a function
$E_{XY}(|\psi\r_{XY})$ which describes  the degree entanglement 
shared between the systems 
$X$ and $Y$. Suppose $n\le m$. 
The most general pure state of the whole system, 
for which the entanglement between the subsystems $X$
and $Y$ is equal to the entanglement shared by pairs of qudits
 is given by the expression 
\be
\label{4.1}
|\Phi\r_{XY}=\left(\bigotimes_{j=1}^n|\chi\r_j\right)\otimes 
|rest\r_{Y_{n+1},\dots Y_m} \, 
\ee
where $|\chi\r_{j}=\sum_k 
\sqrt{a_{k_j}}|\chi_k\r_{X_j}\otimes |\chi_k\r_{Y_j}$ is 
state written in Schmidt basis. The entanglement between the 
systems $X$ and $Y$ which are in the state (\ref{4.1}) is given
can be represented by  the expression
\be
E_{XY}(|\Phi\r_{XY})=\sum_{jk} a_{k_j}\log a_{k_j}.
\label{4.2}
\ee
The equality $E_{XY}(|\psi\r_{XY})=E_{XY}(|\Phi\r_{XY})$
is not sufficient to determine whether
the states $|\psi\r_{XY}$ and $|\Phi\r_{XY}$ are locally unitary equivalent.
Let us rewrite both states in their Schmidt bases
\be
\label{4.3}
|\psi\r_{XY}&=&\sum_{r=0}^{d^n-1}\sqrt{q_r}|\alpha_r\r_X\otimes|\beta_r\r_Y\,
,
\\
\nonumber
|\Phi\r_{XY}&=&\sum_{r=0}^{d^n-1}\sqrt{p_r}|r\r_{X}\otimes
\left(|r\r_{Y_1,\dots Y_n}\otimes |rest\r_{Y_{n+1},\dots,Y_m}\right)\, ,
\ee 
where $d=dim({\cal H}_{A_j})$, $r$ is a base d number which 
represents the state of $n$ qudits, e.g. $r=4$ in a binary form 
represents a state of 5 qubits $|0,0,1,0,0\r$, or in the base 3 form 
it represents a state of 5 qutrits $|0,0,0,1,2\r$.

The eigenvalues of the reduced states are $\{q_r\}$ for the state 
$|\psi\r_{XY}$ and $\{p_r=a_{k_1}\dots a_{k_n}\}$ for the state $|\Phi\r_{XY}$,
where $r=(k_1,\dots,k_n)$ in the number base-$d$ form, i.e. $k_j=0,1,\dots,d-1$.
Theorem 1 implies that these sets of eigenvalues have to be identical in
order to ensure 
 local unitary equivalence. Our aim is to find some condition under which
the equations 
\be
\label{system}
q_r=p_r=p_{(k_1,\dots,k_n)}=a_{k_1}\dots a_{k_n}.
\ee
are satisfied.
In what follows we have $d^n$ equations with $nd$ undetermined 
variables. These equation, in general, do not have a solution.
In that case 
the states are not locally unitary equivalent and the entanglement shared
by systems $X$ and $Y$ cannot be prepared by local unitaries from bi-partite
entanglement shared between the subsystems. 
If the system of equations (\ref{system}) has a solution then the local
unitary transformations are given by relations
\be
U_X|\alpha_r\r_X&:=&|\alpha_r\r_X
\nonumber \\
 U_Y|\beta_r\r_Y &: = &|r\r_{Y_1,\dots,Y_n}\otimes
|rest\r_{Y_{n+1},\dots,Y_m}.
\label{4.4}
\ee
We note, that 
from above,   
the local unitary equivalence
between the $|X\r$ state and $n$ EPR pairs shared between systems $X$ and $Y$
follows (see Section \ref{sec2}).

\section{Conclusion}
\label{sec5}
In this paper we have analyzed the problem of finding under which 
conditions  two states 
states $\varrho$ and $\sigma$ of a multi-partite system
are locally unitary equivalent. We have  presented a set of conditions
which have to be satisfied in order that the two states are locally
unitary equivalent. We have presented a set of parameters which
play the role of invariants under the action of local unitary
transformations.  In addition, 
we have shown how these 
invariants can be used to  characterize the local unitary
equivalence of states. 

We have shown that the problem of local unitary equivalence
for multi-partite systems can be reduced to the problem of analysis
of bi-partite systems. Specifically, 
it is enough to investigate the
case of bi-partite local unitary equivalence and to show sufficient
conditions of this property for non-degenerate density matrices.

We left two questions opened: the local unitary equivalence for 
degenerate density matrices and the characterization of the connection
between  multi-partite and
bipartite entanglement, since, as it is well known that there exists
an intrinsic multi-partite entanglement \cite{Thapliyal99,Wootters2000}.

Finally,  we have analyzed multi-qudit systems.
We have shown, that when analyzing  local unitary
 equivalence of two states it is useful to consider the splitting
of the whole system into two subsystems. 
$X$ and $Y$. In this case a system of equations of the form (\ref{system})
results.
We have shown that these equations might not have a solution from which
it directly follows that the state under consideration 
cannot be prepared by local unitary
transformations from any number of partially entangled pairs
shared between two systems $X$ and $Y$ (i.e. these states cannot be
prepared just using a 
bi-partite entanglement as a resource and local transformations on
subsystems $X$ and $Y$).

\acknowledgements
We thank Mark Hillery for discussions and for reading the manuscript.
V.B. thanks Chris Fuchs for a stimulating discussion.
This work was supported by the IST project EQUIP under the contract
IST-1999-11053 and by the grant GACR 201/01/0413.


\end{multicols}

\begin{thebibliography}{99}

\bibitem{Nielsen2000}
See for example:
M.\ A.\ Nielsen and I.\ L.\ Chuang,
{\em Quantum Computation and Quantum Information}
(Cambridge University Press, Cambridge, 2000).

\bibitem{Gruska1999}
J. Gruska,
{\em Quantum Computing}
(McGraw-Hill, London, 1999).

\bibitem{Peres1995} 
A. Peres, 
{\em Quantum Theory: Concepts and Methods} (Kluwer Academic Publishers, 1995). 


\bibitem{Werner00}
K.G.H.Vollbrecht,R.F.Werner,{\it Entanglement measures under symmetry},
LANL preprint archive {\tt quant-ph/0010095}

\bibitem{Sudbery01}
A.Sudbery,
{\it On local invariants of pure three-qubit states} 
{\em J.Phys.A} {\bf 34}, 643-652 (2001), 
also available at LANL preprint archive {\tt quant-ps/0001116}

\bibitem{Barnum01}
H.Barnum, N.Linden, {\it Monotones and invariants for multi-particle quantum 
states}, LANL preprint archive {\tt quant-ph/0103155}

\bibitem{Stelmachovic01}
P.\v Stelmachovi\v c and V.Bu\v zek, {\it 
Entanglement in Ising model}
(in preparation).

\bibitem{Thapliyal99}
   A.~V.~Thapliyal,
      Phys. Rev. A {\bf 59}, 3336 (1999);
   J.~Kempe,
      Phys. Rev. A {\bf 60}, 910 (1999).     

\bibitem{Wootters2000}
  S.~Hill and W.~K.~Wootters,
       {Phys. Rev. Lett.} {\bf 78}, 5022 (1997);
  W.~K.~Wootters,
        Phys. Rev. Lett. {\bf 80}, 2245 (1998).
 V. Coffman, J. Kundu, and W.~K. Wootters,
        {\tt  quant-ph/9907047} (1999)
  W.~K. Wootters,
        {\tt arXiv quant-ph/0001114} (2000).
                                              

\end{thebibliography}
\end{document}